\documentclass[10pt,a4paper]{article}
\usepackage[english]{babel}
\usepackage{amsfonts}
\usepackage{graphicx}
\usepackage[utf8]{inputenc}
\usepackage{johd}
\usepackage{amsmath}
\title{The Tech Decoupling}

\author{ Monika Baloda \\
        \small{A. Gary Anderson Graduate School of Management, University of California, Riverside} 
}

\date{March 31, 2023} %leave blank

\begin{document}

\maketitle

\begin{abstract} 
\noindent 
Financial market volatility is an important input for investment, option pricing, and financial market regulation and there is no ambiguity in saying that technology is the engine of growth. 
In this project, I investigate the co-movement of technology and non-technology sectors for the last seventeen years. Calling the close co-movement ‘coupling’,  I show that after the year 2015, the technology and non-technology sectors decouples in terms of their levels and volatility. I argue that covid-19 shock is not the reason behind this decoupling phenomenon.
In addition to this, I show that the technology sector is a leading indicator of growth for the rest of the economy.  I use ARIMA modeling and stationarity tests to process our time series for analysis and testing. We find that the technology sector index follows ARIMA(3,1,3), while the non-technology sector follows ARIMA(2,1,4) model.  I discuss the steps of data wrangling, pre-processing, missing value treatment, and exploratory data analysis (EDA). I discuss the merits and shortcomings of my work wherever required to correctly interpret my results. 

\end{abstract}


\section{Introduction}
Two main pillars of the corporate sector are Finance and Technology. Naturally, they are very attractive topics to work on. I, in this paper, dig deeper to find out some insights on the intersection of the two. The main focus of the work is on volatility patterns. I ask questions like: ``does the technology sector more volatile than the rest of the economy?''.  Are there systematic differences in volatility patterns over time in technology and the rest of the economy? I also investigate the inter-relationship between the two series, whether one can be a leading indicator for the other and whether one Granger causes the other. The paper covers the time period from 2006 to 2023,  around seventeen years.  \\[2mm]
\underline{\textit{Motivation}}-- Suppose there is a person A living in a world where the only means of income is a paycheck from the government and daily living expense is \$100. Ms. A is allocated \$36500 per year from the government. There are two possible streams of payments: One is \$100 daily and
the second payment stream is uncertain, she may get \$500 on one day or zero on some other day. Which payment stream will she choose? The obvious answer is that she will choose the first one because, in the second scheme, she will be worse off. In technical terms, this can be said ``Ms. A prefers low volatility over high volatility payment stream''. Several researchers have quantified the consequences of financial market volatility on labor incomes, economic growth, and health outcomes. Therefore, volatility is an important topic to research, it is an important input for investment, option pricing, and financial market regulation. 

There is an influential theory in economics led by Prof. Paul Romer (2018 Nobel laureate in economics) that it is the technology that drives the growth in the economy. If there is no technological growth, then the economy will become stagnant and incomes will no longer increase. Beyond economics, there are in general people interested in knowing how technology is going to affect lives. I, in particular, am fascinated by this theory and have decided to test it (at least partially) in this project. \\[2mm]
Combining these two important pillars: volatility and technology, I decided to investigate the financial market volatility of the technology sector and how it relates to the rest of the economy. 
Therefore, I collected data on the technology sector's index returns in NASDAQ100, and fortunately, the NASDAQ100 gives us the non-technology index as well. These two daily time series spanning time periods from 2006 to 2023 are the backbone of my paper. I follow the famous time series analysis book by James Hamilton (2020) for my work. NASDAQ100's NDXT index is an index of stock prices of NASDAQ100 companies that belong to the technology sector. Let's call this time series $y_t$. Similarly, NASDAQ100's NDXX index is an index of stock prices of NASDAQ100 companies that do not belong to the technology sector. Let's call this series $x_t$. In the first section, I test whether the time series $y_t$ is a leading indicator for the time series $x_t$. Details like how exactly the technology sector is defined are given in later relevant sections.
\\[2mm]
\textit{Concrete Research Questions:}-- My concrete research question is ``Whether the volatility of the technology sector is different from the volatility of the rest of the economy?''. I test the various statistics e.g. mean, quantiles, distribution, etc of the two-time series in order to make a good comparison. To start with I use a simple \textit{ANOVA}-test to compare the means of volatility of two samples. We later disaggregate our data into different time periods in order to test the same volatility relationship in different time periods. We find that indeed the relationship varies in different periods. To complement this analysis, I also check the interrelationship between tech and non-tech series using standard time series tools like \textit{Granger-causality}. I also test whether the technology sector can be a leading indicator for the rest of the economy, i.e. what happens in the technology sector today will happen in other sectors in near future.  The results of statistical tests e.g. whether the technology leads the growth in the rest of the economy can be interpreted differently by different people. Therefore, writing down a concrete research question makes things easy for me to express it to the reader. \\[3mm]
\underline{\textit{Paper Organization:}}--
In section-2, we get to know our time series more closely. This section does data wrangling, exploratory data analysis, and pre-processing of the data for the main analysis. We start by discussing why the time series regime is different, then we talk about data sources and cleaning. The main EDA part is to draw graphs, see summary statistics, and appropriate ARIMA modeling. Later we do stationarity analysis and make our series stationary so that we can analyze them later. 
Section-3 and Section-4 are our main analysis parts. We analyze decoupling in these two sections, first by taking a full sample and then by dividing the data into several intervals. We discuss the takeaways and limitations whenever required. In section-5, we do a complementary analysis to test whether technology leads the rest of the sector. In the later part of section-5, we respond to the possible critique of external validity.  Finally, we conclude.

\section{Data Wrangling and Exploratory Data Analysis (EDA)}

\subsection{What's so Special about Time Series Analysis?} 
There are mainly three types of data sets: cross-sectional, time-series, and panel data. Cross-sectional data are taken on many subjects at one point in time, for example collecting students' GPAs in the Fall-2022 quarter. Time series data is collecting data on a single subject over many (equally spaced) time periods, for example: my GPA in each quarter from starting of school till I graduate. Panel data combines both panel and time series, collecting data on many subjects for multiple time periods. 

Suppose $i$ is a subscript for the subject, it varies from $i=1$ to $i=N$ and $t$ is the subscript for time which goes from $t=1$ to $t=T$. 
Below we write, three regression equations to model a simple linear regression under three different data regimes.
\begin{align*}
    \text{Cross-sectional } & \quad y_i= \beta_0 + \beta_1 x_i +\epsilon_i\\
    \text{Time-series } & \quad y_t= \beta_0 + \beta_1 x_t +\epsilon_t\\
    \text{Panel data} & \quad y_{it}= \beta_0 + \beta_1 x_{it} +\epsilon_{it}
\end{align*}
Our focus in this section of the project is time series. Now, we come to the main question that why time series are so special. The main assumption in cross-sectional and panel data on data is that data is independently and identically distributed (\textit {i.i.d.}). The meaning is that one data point is drawn independently to the other one, similar to the $rnorm()$ function in $R$. However, this assumption is very hard to justify for the time series data. Almost, all the time, time series data is not \textit {i.i.d.}, they show some kind of dependence structure with their previous values. Since this fundamental assumption on data is not satisfied, therefore classical OLS estimation theory becomes invalid for time-series data. Hence, we cannot use the traditional linear regression methods, which is why time series is special. 

Given that we cannot use the classical OLS estimation technique, we need to use some methods which take into account the peculiar problems time series data structure poses. We, therefore, make use of time-series estimation techniques, we refer Hamilton (2020) book for the rest of the discussion on time series models.
\subsection{Data}
Our main goal is to see the relationship between the technology sector and other sectors of the economy. So we collect data on two time series reflecting the same. We describe the source, concrete definitions of the variables used and data cleaning process in this section. 

\subsubsection{Data-source} Our data cover a time period starting from March 21, 2006 to March 21, 2023. The tech index in NASDAQ 100 was constituted for the first time on Feb 22, 2006. Because just after the launch of an index, there may be irrational fluctuations in the index. In order to avoid this, we collect data starting from one month after i.e. March 21, 2006. We stop on March 21, 2023 to complete round-off the time period to 17 years. Our data is collected at a daily frequency. Therefore, we have a total of 4297 data points for a given time series.
We download data from \href{https://finance.yahoo.com/quote/%5ENDXT/history?period1=1142899200&period2=1679356800&interval=1d&filter=history&frequency=1d&includeAdjustedClose=true}{yahoo-finance} website. 

\subsubsection{Variable definitions}
We collect the information on two main time-series : technology and non-technology 
\begin{enumerate}
    \item \textbf{Tech: NASDAQ 100 Technology Sector (NDXT)}: This index reports the equal-weighted index based on the securities of the NASDAQ-100 Index that are classified as Technology according to the Industry Classification Benchmark (ICB) classification system. On February 22, 2006, the NASDAQ-100 Technology Sector Index began with a base value of 1000.00. To know further on how technology sectors are defined, one can refer to the ICB definitions \href{https://www.ftserussell.com/data/industry-classification-benchmark-icb}{here}.
    \item \textbf{Non-tech: NASDAQ 100 Ex Tech Sector (NDXX)}: This index reports an equal-weighted index based on the securities of the NASDAQ-100 Index that are \underline{not} classified as Technology according to the Industry Classification Benchmark (ICB) classification system. Similar to the tech index, this index too began with a base value of 1000.00 on Feb 22, 2006.
\end{enumerate}

\subsubsection{Data cleaning and pre-processing}
Broadly, our data is already cleaned and processed by the data provider's website. 
We hardly have any missing values in our data, therefore no missing value treatment is required. Since the data is taken from the same stock market index of the same country (USA), therefore, the trading days and economic factors are common to both tech and non-tech time series in our case. Therefore, the matching of the same reference unit is not an issue. 

In particular, in the R-markdown notebook, we show that our series $y_t$ has three missing values out of  4277 and the other series $x_t$ does not have any missing value. Therefore, we just drop the three observations as they are not going to make any big difference in our analysis. After this deletion, we are left with  4274 observations or rows in total.

After cleaning the data, we need to process it in $R$ in order to analyze it as a time series object. We mainly use $R$-packages 
\href{https://www.rdocumentation.org/packages/xts/versions/0.13.0/topics/xts}{xts} and \href{https://cran.r-project.org/web/packages/zoo/zoo.pdf}{zoo} for our data-processing and later for analysis. First of all, we convert the data two time-series into numeric form. Then we read the \textit{Date} column as date-object in $R$. After this, we convert the two time-series indices stored in two columns of a data-frame object into a time-series object in $R$. Now $R$-software recognizes our two series as time-series object. Now, we can apply the pre-built functions meant for time series objects on our series. Therefore, we are ready to apply functions to do data exploration and deeper analysis. Hence we move to the exploratory data analysis (EDA) section next.

\subsection{Exploratory Data Analysis (EDA)}
For simplicity from here onward, we call the tech sector index (NDXT) time series $y_t$ and non-tech index (NDXX) time series $x_t$. Before doing any analysis, we study each of the time series in this section. We want to study their properties such as stationarity, auto-regressive order, and moving-average order. 

\subsubsection{Tools}
In this section, we briefly summarize the tools used for data exploration. These tools are typically different from the \textit{i.i.d.} data regime, therefore, I thought that I should give a brief introduction of these tools before using them. 
\begin{enumerate}
   \item \textbf{ACFs: } The autocorrelation function (ACF) is a plot of the correlation between a time series and its lagged values. It is a useful tool for understanding the degree of autocorrelation in a time series and can be used to identify the order of an autoregressive (AR) or moving average (MA) model.
The ACF plot here is showing lags on the x-axis and correlation coefficients on the y-axis. The correlation coefficients range from -1 to 1, with 1 indicating a perfect positive correlation and -1 indicating a perfect negative correlation. A value of 0 indicates no correlation.

    \item \textbf{PACFs: }In James Hamilton's book ``Time Series Analysis", the PACF is introduced in Chapter 2 as a method for determining the order of an autoregressive (AR) model. The PACF is used to identify the number of lags that need to be included in the AR model by examining the magnitude and significance of the partial autocorrelations at each lag. A large partial autocorrelation at a particular lag indicates a strong correlation between the original series and that lagged value, which suggests that the lag should be included in the AR model.
Hamilton notes that both ACF and PACF are useful tools for identifying the order of an AR model, but they should not be used in isolation. Both of them needed together to decide on AR and MA lag models.
    \item \textbf{Co-integration test: } In order to get an idea about the correlation between two series, we do a naive correlation. To comment on the long-run relationship and with technical confidence, I then do Johanson's co-integration test.

    \item \textbf{Comparative Graph: } We plot the two series together to see how they behave. This gives us an idea of what exactly different thing happens to the two series at some point in time. I didn't plot both series individually to avoid repeated grapes and to save some space. 
    \item \textbf{Summary statistics: } We draw a table to present the five-point summary, average value, the overall change from the beginning date to the last date, and variation in the data. This table gives us an idea about the distribution and overall percentage returns of the series. 

\end{enumerate}

\subsubsection{The Tech Index ($y_t$) Time Series}
In this section, we'll get to know our tech index more in detail. First of all, we want to say that we have data on this index since the time it came into existence in 2006. The base value of the index is 1000. 
We present a table showing a five-point summary and the average value of $y_t$ below in table-\ref{tab:summary_y}
\begin{table}[htbp]
    \centering
     \caption{Summary statistics of tech-index series ($y_t$)} 
    \begin{tabular}{lr | lr} \hline
         Statistics &    Value &Statistics &    Value \\  \hline   
   Minimum   & 525.90 & Mean   &3029.00  \\  
    1st Quartile &1194.20 &Latest Value & 6783.62  \\  
   Median &2138.40 &Starting Value& 1021.11  \\ 
     3rd Quartile &4333.80  &Standard Deviation& 2330.62  \\ 
   Maximum   &9855.40 &Base-value &1000.00 \\ 
    \hline
    \end{tabular}
  \label{tab:summary_y}
\end{table}
The starting value means the value of the index when we started collecting data which is the value on March 21, 2006. The latest value of the index (on March 20, 2023) is 6783, which implies that the tech index has increased by around 6-fold in the last 17 years. This is an impressive annual return of around 33\% on average. We also see substantial volatility in this series with the standard deviation being almost the same order as the average value. \\[4mm] 
\begin{figure}[htbp]
    \centering
    \caption{autocorrelation function (ACF) and Partial ACFs of Tech index (series $y_t$)}
    \includegraphics[width=100mm]{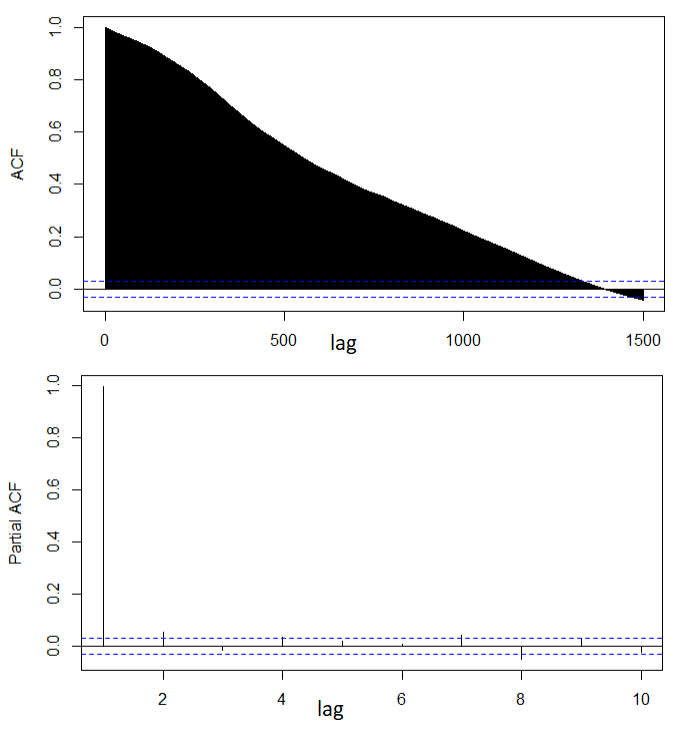}
     \label{ACF_y}
\end{figure}
In order to know more about the time-dependence structure of the series, we plot the autocorrelation function (ACF) and partial auto-correlation function (PACFs) in figure-\ref{ACF_y}. 
In the case of our series $y_t$, the ACFs don't become zero. To be technically correct, ACFs fall in a confidence interval around zero that is not significantly different from zero after taking approximately 1400 lags into account. However, handling such a high number of lags is neither easy nor parsimonious from the modeling point of view. 
Since the autocorrelation function decays very very slowly, there may be a unit root component that is making the series non-stationary for practical purposes. Therefore, we need to do a unit-root test for this series before proceeding further.  We are not deciding on the AR order of the series at this moment. Doing unit-root tests is beyond the capacity of EDA section, 
we defer it to the analysis section to be done later on.

\subsubsection{The Non-Tech Index ($x_t$) Time Series}
In this section, we are going to learn about our non-technology index i.e. our series $x_t$ in detail. The non-technology index is formally known as the Ex-Technology sector index at NASDAQ100 and is represented by NDXX as discussed in the data section previously.  Below, we present a summary table (Table-\ref{tab:summary_x}) that gives us a snapshot of how our series has been in the last 17 years of its existence. 
\begin{table}[htbp]
    \centering
     \caption{Summary statistics of Non-tech index series ($x_t$)} 
    \begin{tabular}{lr | lr} 
    \hline
   Statistics &    Value &Statistics &    Value \\  
   \hline   
   Minimum   & 561 & Mean   &1319  \\  
    1st Quartile &1191 &Latest Value & 4675  \\  
   Median &2177 &Starting Value& 1027  \\ 
     3rd Quartile &3055  &Standard Deviation& 1283  \\ 
   Maximum   &5360 &Base-value &1000 \\ 
    \hline
    \end{tabular}
    \label{tab:summary_x}
\end{table}    
We note that the base value of this index is 1000 but we started collecting data one month after, therefore, our starting value means the value of the index on March 21, 2006. We note that the latest value of the index is 4675, which is a 3.55-fold increase from its starting value. If we convert this change into an average annual return, we find that we get around 21\% annual returns. We note that this return is lower than the technology index return which was 33\% annually. We further note that the volatility of this series is less than $y_t$ because the standard deviation here is half of the average value while in the series $y_t$, it was around 75\% of the mean value. \\[4mm]
In order to uncover the time-dependence pattern of this series, we look at the ACFs and PACFs.
\begin{figure}
\caption{autocorrelation function (ACF) and Partial ACFs of Non-tech index (series $x_t$)}
    \centering
    \includegraphics[width=100mm]{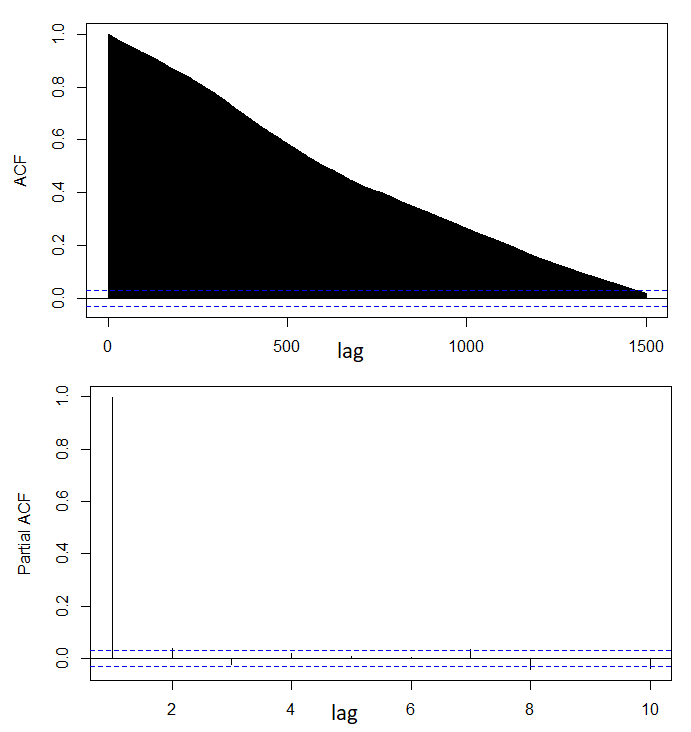}
    \label{ACF_x}
\end{figure}
The two graphs i.e. ACFs and PACFs of this series (Figure-\ref{ACF_x}) are qualitatively similar to that of series $y_t$. We note that the ACFs of this series decay even slower than series $y_t$. The PACFs go to zero immediately in the first lag. Therefore, the conclusions are similar to that of what we said about series $y_t$. We need to do a unit root test to first check whether the series is 
stationary or not. Therefore, we defer the remaining analysis for the later sections.

\subsubsection{Analyzing both time series together}
\begin{figure}[htbp]
    \centering
     \caption{Plot of Tech (series $y$) and Non-tech (series $y$) indices over time}
    \includegraphics[width=160mm]{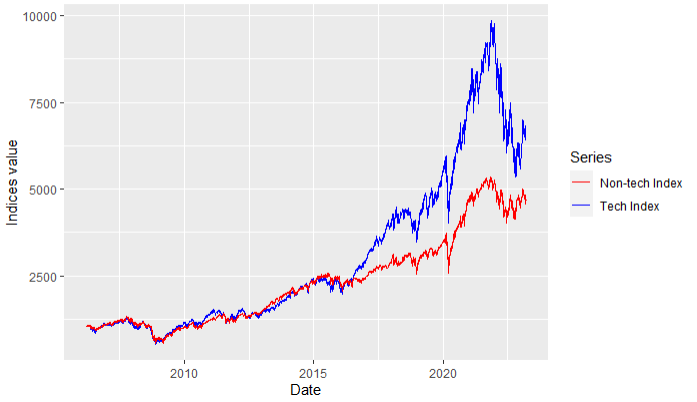}
     \label{graph_xy}
\end{figure}

\begin{figure}
    \centering
    \caption{Cross-correlation between Tech ($y$) and Non-tech($x$) indices upto 1000 lag}
    \includegraphics{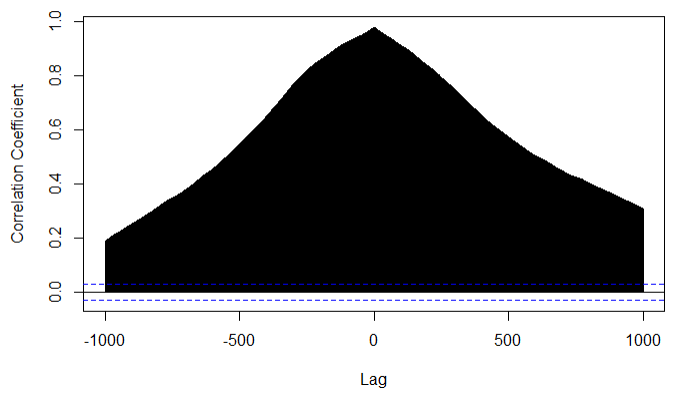}
    \label{ccf_xy}
\end{figure}

Analyzing the two series together gives very important insights into their relationships. We report two graphs and one test in this section. First of all, we simply plot a graph of two-time series in a single plot (Figure-\ref{graph_xy}). \\[2mm]
We can see that the two series closely mingle with each other till around 2016. After that, they started de-coupling. The covid-19 shocks further widened this decoupling of the two series. Too much close co-movement of the two series directly tells us that these series may be closely correlated.\\[2mm]
In order to see their cross-correlations, we draw a graph (Figure-\ref{ccf_xy}). We keep the maximum lag order equal to 1000, meaning that the cross-correlation is to be calculated between the current value and the 1000th lagged value of the two series. 
We note that the correlation coefficient decays very very slowly like the ACFs of individual series. This graph suggests that what happened 1000 trading days ago in the tech index will affect the non-tech index today. This claim looks to be easily believable. There could be some spurious correlation. We need to test the long-run relationship between $x_t$ and $y_t$ series. 

A spurious correlation is a concern when analyzing time series data because many economic and financial variables are highly correlated with each other, even when they do not have a causal relationship. This can be due to the influence of common external factors or to the fact that the variables are non-stationary and exhibit trends that are not related to each other. Co-integration, on the other hand, is an important concept in time series analysis because it allows for the identification of long-term relationships between non-stationary variables. Hamilton emphasizes the importance of testing for co-integration before estimating regression models, as failure to do so can result in spurious regression and incorrect inference. \\[2mm]
Hamilton (2020) also discusses various methods for testing for co-integration, including the Engle-Granger two-step procedure and the Johansen procedure. These methods involve testing for the presence of a common stochastic trend between variables and can help to identify whether a meaningful long-run relationship exists between variables. Therefore, we go ahead and do a co-integration test here. Here, we only report (Table-\ref{tab:johanson}) the test statistics and critical values which are necessary to take a decision. The rest of the details of the result of  Johansen's test are omitted for the simplicity of the presentation. 
The Johansen cointegration test is implemented by the \textit{ca.jo()} function in R. The null hypothesis is:
\begin{align*}
    H_0 : & \text{ There is no co-integration among the variables in the system}\\
\end{align*}
In other words, the null hypothesis is that there are no linear combinations of the variables that are stationary and share a long-run relationship.
\begin{table}[htbp]
    \centering
\caption{Values of Johansen's Co-integration test statistic and critical values of test}
    \begin{tabular}{cccccc}
    \hline
order & test-statistics & 10\% critical value & 5\% &1\%\\
\hline
$r <= 1$ &  0.0 & 6.50&  8.18& 11.65\\
$r = 0 $ &  3.6& 15.66 & 17.95 &23.52\\
    \hline 
    \end{tabular}
    \label{tab:johanson}
\end{table}\\[2mm]
\textit{Interpretation of the results}: In order to conduct a co-integration test, the two series have to be stationary. But we are not sure in our case whether they are stationary or not. \underline{Assuming that both $y_t$ and $x_t$ are stationary}, then this test statistic (=3.6) is less than the critical values at 1\%, 5\%, and 10\% significance levels implying that we cannot reject the null hypothesis. That is, there is no cointegration. Hence, we conclude that there is no evidence of the presence of co-integration or long-run relationship between series $y_t$ and series $x_t$. Since there is no co-integration, we cannot say that there is a long-run relationship between our two series. But the cross-correlation graphs tell the opposite story. However all this is under the assumption that both of our series is stationary, since we haven't tested this assumption, our conclusion may be wrong. Therefore, we need to do a detailed stationarity analysis before concluding anything. 

\subsubsection{Takeaways from EDA} In this section we got to know our time series better. We now know how they have behaved in the last 17 years in terms of volatility, and distributions. We also looked at the cross-relationship between the two variables. Our main takeaway is that there is a strong possibility that both time series are non-stationary. Using cross-correlation and co-integration tests, we further strengthen the requirement of stationarity analysis. Note that the results of our co-integration test may not be valid because we don't know whether $x_t$ and $y_t$ are co-integrated or not.

\subsection{Stationarity Analysis}
Many of the fundamental results and techniques in the field rely on the assumption of stationarity, so violating this assumption can make it difficult to apply these methods correctly. Ensuring that a time series is stationary is a crucial step in the analysis of time series data, and can help to avoid common pitfalls and errors in modeling and inference.

In the EDA section, we have seen that there is a strong need for us to know whether our time series ($y_t$ and $x_t$) are stationary or not. Without this knowledge, all of the analysis that we are going to do further may not be correct. In this section, we first perform the tests to check whether the time series is stationary or not. If we found them to be non-stationary then, we transform them into stationary time series by the appropriate method. We do detrending if there is a deterministic time trend present. On the other hand, if there is a stochastic trend, then we do differencing.  

\subsubsection{What is a Stationarity in Time Series}
In simple words, a stationary time series is a series whose statistical properties such as mean, variance, and co-variance remain the same over time. There is more than one notion of stationarity, like strong stationarity, weak or covariance stationarity etc. However, weak stationarity is the main notion used in the literature. We define it in the following text. We call a time series $y_t$ to be weak or covariance stationary if the following holds: 
\begin{enumerate}
    \item $\mathbb{E}(y_t)=\mu $ :  The mean of the series doesn't depend on time. 
    \item $Var(y_t) = \sigma^2_y <\infty \quad$: Variance of the series is finite and doesn't depend on time. 
    \item $Cov(y_t, y_{t-s})=\gamma(s) \quad$: Co-variance between any two-time terms depends only on the time difference between the point and not on the absolute time.
\end{enumerate}
So basically the three properties are not a function of time. In other words, we can say that $\mathbb{E}(y_t)=\mathbb{E}(y_{t+i})=\mu$ for all $i$. Also, $Cov(y_1, y_2)=Cov(y_2, y_3)=Cov(y_{99}, y_{100})$.

\subsubsection{Why do we care about Stationarity?}
Having defined the stationarity, I now discuss, why is this a good property and why care about it. There are several reasons but I want to highlight three main reasons. First, 
stationarity makes the behavior of the series consistent and predictable, making it easier to model and analyze.
Second, is that many of the standard time series models, such as autoregressive integrated moving average (ARIMA) models, assume stationarity. These models are widely used in forecasting and other applications, so it is important to ensure that the underlying data is stationary before applying them.
The third reason why stationarity is important is that non-stationary time series can exhibit spurious correlations, which can lead to misleading conclusions. For example, if two non-stationary time series appear to be highly correlated, this may simply be due to the fact that both series are trending upwards over time, rather than a genuine relationship between the two series.

\subsubsection{Testing for Stationarity}
Consider the following simple AR(1) time series model:
\begin{equation*}
    y_t= \alpha + \beta_1 y_{t-1} + \epsilon_t 
\end{equation*}
The coefficient $\beta_1$ represents the marginal effect of $y_{t-1}$ on $y_t$.
If $ | \beta_1 |<1 $, then the effect of \textit{lag} term on the current time series dies out with time, and therefore, we call this series to be stationary. But if this condition is not satisfied then the series is said to be non-stationary. 
If $ | \beta_1 | = 1 $, then the series is non-stationary and we say that there exists a unit root. Our goal in this section is to know whether this beta is significantly different from unity or not.  This was a simple example of an AR(1) model, a real-life time series model may be more complex than this. In our testing of unit root, we do not assume that our series follows AR(1) model, the discussion here is for exposition purposes only. \\[2mm]
We use the Augmented Dickey-Fuller (ADF) test to test for unit roots in both of our series. The null hypothesis of the ADF test is : 
\begin{align*}
H_0 :& \text{ Series has a unit root hence is non-stationary}\\
H_1 :& \text{ Series is stationary}
\end{align*}
Therefore, if we find \textit{p-value} to be greater than 0.05, then we cannot reject the null hypothesis at a 95\% level of significance. It will mean that our series has a unit root and hence is not stationary. \begin{table}[htbp]
    \centering
    \caption{Result of ADF-test for Stationarity}
    \begin{tabular}{lcc}
    \hline 
    Time-series & Test-Statistics & p-value \\
    \hline 
      Tech($y_t$)       & -2.11     &0.53 \\
      Non-tech ($x_t$)  &-2.48      &0.37 \\
      \hline    
    \end{tabular}
    \label{tab:ADF}
\end{table}
The result shows that the \textit{p-value} is greater than 0.05, therefore, we cannot reject the null hypothesis that both series are non-stationary. Therefore, both of our series are not stationary. Hence, the correlation that we saw in EDA section cannot be relied upon. 

\subsubsection{Transforming Our Series into Stationary Series}
We generate a first-differenced time series in order to do away with unit root in the original series. The first-differenced series ($\Delta y_t$) looks is generated by taking a difference of one period lag from the current period value. That is: 
$$
 \Delta y_t = y_{t}-y_{t-1}   \quad  \quad \quad  \quad \text{and} \quad  \quad  \quad  \quad \Delta x_t = x_{t}-x_{t-1} 
$$
Because the first-differenced time series subtracts the one-period lag, there is no $\Delta y_1$ exists because we don't have $y_0$. Therefore, one missing value will be introduced when we take the first difference. Therefore, we need to delete one observation in order to do further analysis. After doing so, we test whether the transformed series by first-differencing is stationary or not. \\[2mm]
We again perform ADF test on the transformed series to check their stationarity. The results are reported below (Table-\ref{tab:ADF_fd} ):
\begin{table}[htbp]
    \centering
    \caption{Result of ADF-test for Stationarity on Transformed Series}
    \begin{tabular}{lcc}
    \hline 
    Time-series & Test-Statistics & p-value \\
    \hline 
     First-differenced Tech($\Delta y_t$)       & -16.824    & $<0.01$ \\
     First-differenced Non-tech ($\Delta x_t$)  & -16.392    & $<0.01$ \\
      \hline    
    \end{tabular}
    \label{tab:ADF_fd}
\end{table}
The interpretation of the results is that we note that both of our first-differenced series $\Delta y_t$ and $\Delta x_t$ are stationary at 99\% confidence level. In other words, with more than 99\% confidence, we can say that the first-differenced series $\Delta y_t$ is stationary. The same line applies for $\Delta x_t$. Now since our series are stationary, we can apply the usual time series models. 

\subsubsection{Appropriate Time Series Modelling : ARIMA Model}
Knowledge of the modeling structure of a series always helps us in our analysis. It is like approaching something with some information about it versus just hitting the target in the dark. Therefore, we go ahead and determine the order of auto-regression and moving average in our transformed time series. \\[2mm]
To determine the order of the transformed series, we use the in-built function in $R$. The function that directly gives the ARIMA order a series is \textit{auto.arima}(), one can see the detailed documentation of the method on the official website of R by clicking \href{https://www.rdocumentation.org/packages/stats/versions/3.6.2/topics/arima}{here}. We employ \textit{auto.arima} function to get the AR and MA orders of the transformed and original series. We tabulate the results in Table-\ref{auto_arima}
\begin{table}[htbp]
    \centering
    \caption{ARIMA Order of the Time Series using \textit{auto.arima} function of R}
    \begin{tabular}{lccccc}
    \hline 
    Time-series & AR-order &I-order & MA-order \\
    \hline 
    Tech ($y_t$)       & 3&1 &3  \\
    Non-tech ($x_t$)  &2 & 1&4 \\
     First-differenced Tech ($\Delta y_t$)       &3 &0 &3 \\
     First-differenced Non-tech ($\Delta x_t$)  &2 &0 &2 \\
      \hline    
    \end{tabular}
    \label{auto_arima}
\end{table}\\[2mm]
In the table-\ref{auto_arima}, AR-order means up to how many lag terms of itself a series depends on. The \textit{I}-order means the integration order i.e. it counts a number of unit roots in the time series. If the number or unit roots or I-order is zero, then our series is said to be stationary. The MA order tells us about the structure of the error term. An MA-order being equal to zero means the error term is white noise. \\[2mm]
We note that the I-order of the original $y_t$ and $x_t$ series is one 
for the transformed series, it is zero. Therefore, we can confirm our earlier findings that both the original series have one unit root each. Confirmation of our findings through an alternate method gives more confidence in our results.
\subsubsection{Takeaways from Stationarity Analysis}
We start this section by discussing why stationarity is important and its definition. We then describe how to test for stationarity. Implementing the stationarity test, we learned that both of our series i.e. Tech index and Non-tech index are non-stationary. In the following text, we transform our series to make it stationary. We note that both series become stationary after taking the first difference. We then confirm this finding by an alternate in-built function in R. 

Finally, we close this section by noting that the Tech index ($y_t$) follows ARIMA(3,1,3) model, and the Non-tech index ($x_t$) follows ARIMA(2,1,4) model.

\subsection{Takeaways from this Exploratory Data Analysis}
This exploratory time series analysis gave us great insights. I realized how much pre-processing is needed before jumping to the main model. I summarize my takeaways in the following points:
\begin{enumerate}
    \item Technology index series ($y_t$) follows ARIMA(3,1,3) model. On the other hand, the Non-technology index follows an ARIMA(2,1,4) model. The I=1 order implies that both the series are non-stationary but their first differenced series becomes stationary.      
    \item There is a very close relationship between tech ($y_t$) and non-tech ($x_t$) indices, however, we observed a decoupling after 2016 which further widened with covid shock in 2019. 
      \item There seems to be a structural break in the tech index series in 2016 which further widens due to covid shock in 2020.\\[2mm]
\end{enumerate}
This section paves a path for further analysis. I analyze the volatility patterns in greater detail in the next section. I also do some complementary analysis to support the main analysis.

\section{The Tech Decoupling in Levels}
As discussed in the previous sections, the technology and non-technology indexes we are using came into existence for the first time in 2006. Both of them started from the same base value of 1000. In this section, we are tracing their evolution over time. In particular, we want to check whether there exists a particular point in time or period in which these two-time series are decoupled from each other.\\[2mm]
We have already seen that both technology and non-technology index series are non-stationary having one unit root. Therefore, we make it a stationary series transforming by first differencing. We use this transformed series for the analysis in this section.  

\subsection{Testing Decoupling in Levels Using t-test in Full Sample}
\subsubsection{Testing Procedure}
By testing coupling we mean whether the two series are close enough to each other. We call them decoupled if we can find statistically significant differences between their level in a given period of time. 
Here we employ a basic t-test to check whether the mean of the two indices is significantly different from each other. Suppose the mean of the tech index is $\mu_y$ and the mean of the non-tech index is $\mu_x$. Then our null hypothesis is $\mu_y = \mu_x$. To make it a simple t-test, we take the difference of the two 
series and test the mean equal to zero. That is our hypothesis is: 
\begin{align*}
    H_0 &\quad \mu_y - \mu_x=0\\
    H_1 &\quad \mu_y - \mu_x \neq 0
\end{align*}
We test the null hypothesis of the difference of the two series equal to zero against the alternate of it being non-zero. If we can reject the null hypothesis then we will able to say that the two series are not coupled in levels. 
\paragraph{Results and Inferences}
We report the t-statistics and corresponding p-value in the table-\ref{fd_mean_test}. 
\begin{table}[htbp]
    \centering
    \caption{Testing for Equality of Levels of First-differenced Time Series (Full-Sample)}
    \begin{tabular}{lccccc}
    \hline 
    Time-series & t-statistics & p-value \\
    \hline 
     $\Delta y - \Delta x$   & 0.6951&0.487  \\
      \hline    
    \end{tabular}
    \label{fd_mean_test}
\end{table}
The results show that we cannot reject the null hypothesis when we consider the whole sample. The results give evidence in favor of the null hypothesis which basically says that overall levels of the tech index and non-tech index are not significantly different from each other. 
\subsection{Testing   Decoupling in Levels in Decomposed Time-Intervals}
In this part, we divide the whole time span of our data into five parts. We tried to keep them of equal size whenever possible. Below, we describe the division into different time intervals and then report the results.
\subsubsection{Division in Multiple Time Periods}
In this section, we split the dataset into different parts. We split our data into four parts covering different data points. The first period covers observations from March 2006 to Feb 2010, the second division covers March 2010 to Feb 2015, the third division covers April 2015 to March 12, 2020, and the last part covers the time after January 20, 2020, to March 2023. We choose a specific date for partitioning the data into third and fourth parts because, as per Center for Disease Center \href{https://www.cdc.gov/museum/timeline/covid19.html#:~:text=January%2020%2C%202020,respond%20to%20the%20emerging%20outbreak.}{website}, the first case of covid in the United States came on this date. We choose this date for splitting the data into the third and fourth parts. We use March 12, 2020 date to do a sensitivity analysis. March 12, 2020, is the date when the first ever stay-at-home order came into existence in the United States. 
After splitting the data into four parts, we do a t-test to compare the mean of volatility in two series.  
We further divide the fourth division into two parts: We call period from Jan 20, 2020 to March 2022 to be covid-time. After March 2022, the third wave in the US got over [\href{https://www.cdc.gov/coronavirus/2019-ncov/covid-data/covidview/index.html
}{CDC ref}], so we call the period after this time to be post-covid.

\subsubsection{Results and Inferences}
We use the stationary transformation of tech and non-tech index and then take their difference to test for coupling. The following table-\ref{fd_mean_test_multiple_period} reports the results of hypothesis testing for multiple time periods. 
\begin{table}[htbp]
    \centering
    \caption{Testing for Equality of Levels of Time Series (Decompsed Intervals)}
    \begin{tabular}{lccccc}
    \hline \hline
    Time-period & t-statistics & p-value \\
    \hline \hline
     Global Financial Crisis (GFC): March 2006 to Feb 2010 &0.366 &  0.715 \\
     Post-GFC: March 2010 to Feb 2015   &  -0.452 & 0.651  \\
     Pre-covid: March 2015 to Jan 20, 2020   &2.299  & \textit{0.023}  \\
     Covid: Jan 20, 2020 to March 2022   &0.567 &  0.571 \\
     Post-Covid: March 2022 to March 2023   & -0.837 & 0.403  \\
      \hline    
      & &  \\
      Division into Two Halves  & &  \\
        \hline
     First Half: March 2006 to Feb 2015 & -0.135 &  0.893  \\
     Second Half: March 2015 to March 2023  &0.718 & 0.473  \\
      \hline \hline 
    \end{tabular}
    \label{fd_mean_test_multiple_period}
\end{table}\\[1mm]
The inferences from these results can be written in the following bullet points: 
\begin{itemize}
    \item The tech and non-tech indices are coupled during 2006-2015. The global financial crisis couldn't break this coupling. 
    \item The pre-covid period (2015-2020) decoupled the tech and non-tech sectors of the economy. The index levels of the two sectors are significantly different from each other at 95\% confidence level. In particular, at 95\% confidence, we can say that the tech index level is higher than the non-tech index during the pre-covid period.  
    \item Contrary to popular belief, the covid-19 period does not contribute to decoupling. 
\end{itemize}

\subsection{Takeaways}
The technology and non-technology sectors are closely coupled for the majority of the time period. If we do a statistical test of the difference in mean level, then we cannot reject the null that the two series are close in the full sample. But if we decompose the sample into several parts, we can see the decoupling of the two sectors. 
Pre-covid period 2015-2020 is the time when the decoupling of technology and non-technology started appearing. Both covid-19 and the global financial crisis do not appear to have a big enough effect on the close co-movement of the tech and non-tech series.

\section{Decoupling in Volatility}
This section is the main heart of the paper. I first discuss the construction of volatility measures. Then I analyze the time series properties of it. We then process the time series to make it stationary so that we can apply the standard time series tests.  
\subsection{Volatility Series:  Construction, Modelling, and Pre-processing}

\subsubsection{Defining Volatility}
Volatility is a measure of ups and downs in a time series. In other words, the spread of a series could be a good measure of ups and down. One such measure is the standard deviation. We use the rolling window method to determine a series's standard deviation and call it a volatility measure. This measure is a standard practice in the literature. The following steps describe the construction of volatility: 
\begin{itemize}
    \item \underline{Step-1: Choose a window size}-- The first step is to choose a window size, which determines the number of historical returns used to calculate the volatility measure. We use the size of the window equal to 100. This is 100 trading days, which is approximately 4 months. 
   \item \underline{Step-2: Calculate returns}-- Next, calculate the returns for the asset over the window period. The return is typically calculated as the percentage change in price over the window period.

   \item \underline{Step-3: Calculate standard deviation}-- Calculate the standard deviation of the returns over the window period. This is the volatility measure for that period.

    \item \underline{Step-4: Shift the window}-- Move the window forward by one day and repeat steps 2 and 3. This will give you a new volatility measure for the next window period.

    \item \underline{Step-5: Repeat}-- Continue shifting the window forward and recalculating the volatility measure until you have covered the entire time period of interest.
\end{itemize}
Our choice of window size is an arbitrary one. Generally, a longer rolling window size will provide a more stable and smoother volatility estimate, but at the cost of potentially missing short-term changes in volatility. On the other hand, a shorter rolling window size will be more responsive to short-term changes in volatility, but at the cost of increased noise and potential instability in the volatility estimate. Since our analysis is long-term (17 years), having a rolling window of 100 trading days which is around four months in a calendar sense is justified. We later do a sensitivity analysis by comparing the volatility measures produced by different window sizes to compare qualitative results.\\[2mm]
Since we are using a window size equal to 100, we lose 100 data points because there are no previous 100 values available for initial observations to form a window, therefore no volatility measure is constructed corresponding to those data points. However, note that there is one volatility data point for each of the remaining observations. Hence, the total number of volatility series will be equal to 100 less than the number of observations in the original series. For the rest of the discussion, we call the volatility series of the technology index \textit{vol\_y} and the volatility series of the rest of the economy index \textit{vol\_x}.

\subsubsection{Time Series Modelling for Volatility}
In our discussion on stationarity, we learned the importance of having a stationary series. Basically, in order to apply the majority of the time series functions, we need to transform the volatility series constructed into a stationary series. Before jumping into whether the volatility series is stationary or not, we plot both volatility series on the time axis. 
\begin{figure}
    \centering
    \caption{Volatility of Tech ($y_t$) and Non-tech ($x_t$) Series} 
    \includegraphics{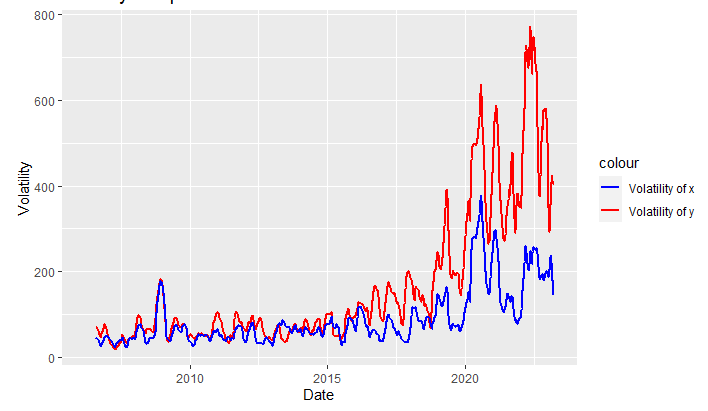}
    \label{fig:vol_xy}
\end{figure}\\[1mm]
From the inspection, the series does not appear to be stationary. To formally confirm this guess, we do a statistical test. 
In the table-\ref{auto_arima_vol}, we describe the appropriate time series model for the two volatility series. 
\begin{table}[htbp]
    \centering
    \caption{ARIMA Order of the Volatility Time Series using \textit{auto.arima} function of R}
    \begin{tabular}{lccccc}
    \hline 
    Time-series & AR-order &I-order & MA-order \\
    \hline 
    Volatility of Tech Index (vol\_y)       & 1&1 &2  \\
    Volatility of Non-Tech Index (vol\_x) &3 & 1&3 \\
      \hline    
    \end{tabular}
    \label{auto_arima_vol}
\end{table}\\[2mm]
We note that both the volatility series are non-stationary. They are integrated of order one. We close this section by saying that the volatility of tech index vol\_y follows the ARIMA(1,1,2) model and the volatility of non-tech index vol\_x follows ARIMA(3,1,3)

\subsubsection{Making Volatility Series Stationary}
In the previous sub-section, we noted that both the volatility series have one unit root each. Therefore, a straightforward step to make it stationary is to take the first difference. We have already done so in the EDA section when we were dealing with series $y_t$ and $x_t$.  
\begin{table}[htbp]
    \centering
    \caption{ARIMA Order of the First-differenced Volatility Time Series}
    \begin{tabular}{lccccc}
    \hline 
    Time-series & AR-order &I-order & MA-order \\
    \hline 
     First-differenced vol\_y ($\Delta$vol\_y)       &1 &0 &2 \\
     First-differenced vol\_x ($\Delta$vol\_x)  &3 &0 &3 \\
      \hline    
    \end{tabular}
    \label{auto_arima_vol_fd}
\end{table}\\[2mm]
After taking the first difference of the volatility series, we do an Augmented Dickey-Fuller (ADF) test to check whether the transformed series is now stationary. For both the series, at 99\% confidence, we can reject the null hypothesis that series are non-stationary. Therefore, the transformed series are stationary. 
\subsection{A Simple t-test on Full Sample}
\subsubsection{Testing Procedure}
A simple $t$-test compares the means of two samples. Suppose the mean volatility of series vol\_y is $\mu_{vol\_y}$ and the mean volatility of series is vol\_y is $\mu_{vol\_y}$. The hypothesis of our t-test is: 
\begin{align*}
    H_0 :  &\quad \mu_{vol\_y} = \mu_{vol\_x} \\
    H_1 :  &\quad \mu_{vol\_y} \neq \mu_{vol\_x}
\end{align*}
An alternate way to test this hypothesis is to just take the difference between the two series and test the mean of the new series equal to zero. That is, our null hypothesis is: 
\begin{align*}
    H_0 :  &\quad  \mu_{vol\_diff} = \mu_{vol\_y} - \mu_{vol\_x}=0 \\
\end{align*}
Since the assumption of the t-test requires us to have data be stationary, we used the first-differenced volatility series. It is because the original volatility series is non-stationary which doesn't satisfy the assumptions of t-test, therefore may yield wrong results.  Therefore by vol\_y we mean the first differenced vol\_y in our testing. 
\subsubsection{Results and Inferences}
We report the t-statistics and corresponding p-value in the table-\ref{vol_fd_mean_test}. 
\begin{table}[htbp]
    \centering
    \caption{Testing for Equality of Mean of First-differenced Volatility Time Series (Full-Sample)}
    \begin{tabular}{lccccc}
    \hline 
    Time-series & t-statistics & p-value \\
    \hline 
     First-differenced vol\_y -  First-differenced vol\_x       &1.551 &0.1209  \\
      \hline    
    \end{tabular}
    \label{vol_fd_mean_test}
\end{table}
We observe that we cannot reject the null hypothesis that the difference of means of two volatility series is different from zero at a 95\% confidence level. Therefore we conclude that in the full sample, the volatility of tech and non-tech indices are not significantly different. 
\subsection{Testing in Decomposed Time Periods}
We use the same time intervals discussed previously. In this section, we report the result for volatility. 
We use the first differenced transformed series of volatility for testing as we did in the previous simple testing for the full sample. For the sake of simplicity, we omit further detail on it here. 
\begin{table}[htbp]
    \centering
    \caption{Testing for Equality of Mean of First-differenced Volatility Time Series (Dis-aggregated Period)}
    \begin{tabular}{lccccc}
    \hline \hline
    Time-period & t-statistics & p-value \\
    \hline \hline
     Global Financial Crisis (GFC): March 2006 to Feb 2010 &-0.765 & 0.4446  \\
     Post-GFC: March 2010 to Feb 2015   &-0.105 &0.9167  \\
     Pre-covid: March 2015 to Jan 20, 2020   &2.623 & \textit{0.0088}  \\
     Covid: Jan 20, 2020 to March 2022   &3.674 & \textit{0.0003}  \\
     Post-Covid: March 2022 to March 2023   &-2.050 & \textit{0.0414}  \\
      \hline    
      & &  \\
      Division into Two Halves  & &  \\
        \hline
     First Half: March 2006 to Feb 2015 & -0.507 & 0.6121  \\
     Second Half: March 2015 to March 2023  &1.686 & 0.0919  \\
      \hline \hline 
    \end{tabular}
    \label{vol_fd_mean_test_multiple_period}
\end{table}\\[1mm]
We observe interesting inferences from the testing in multiple periods. We use the following bullet points to present our findings: 
\begin{enumerate}
    \item We note that there is no difference in volatility of tech and non-tech indices if we see all 17 years together. But there are drastically 
     heterogeneous results for different time periods if we look at more granular levels.  
    \item During the period of 2015 to 2020, the volatility of the technology sector is significantly higher than the volatility of the rest of the economy. We can say that the technology sector decouples with the rest of the economy in terms of volatility in this particular time period.  
    \item We note that the covid-19 period magnifies the decoupling by a wide margin. 
    \item However, the post-covid period somewhat close this gap in volatility by increasing the volatility of the rest of the economy and not by decreasing the volatility of the technology series.    
    \item If we divide the entire sample into two samples, then the latter half is characterized by higher-tech volatility than the rest of the economy with more than 90\% confidence. 
\end{enumerate}

\subsection{Takeaways from Volatility Analysis}
To be very brief in summarizing the takeaways, I note the following bullet points: 
\begin{itemize}
    \item There is significantly higher volatility in the technology sector index than in the rest of the economy after 2015. We call this volatility decoupling of the technology sector from the rest of the economy.

    \item The \textit{tech decoupling} we are talking about is not originated from covid pandemic effect. However, we note that Covid-19 magnifies the decoupling. 
    \item We further note that post covid period close the volatility gap by increasing the volatility of the non-technology sectors. 
    \item Overall, the period after 2015 is characterized by higher volatility, especially led by technology. Non-technology sector later joins the trend. 
\end{itemize}

\section{Complementary Analysis}
\subsection{Does Technology Lead the Growth?}
There is an influential theory in economics that says that technology leads the 
growth in the economy. Prominent research papers include the 2018 Nobel laureate in economics Paul Romer's work in this area. In this section, we aim to test this theory using time series modeling. To practically implement this test, we define the technology to be the stock returns of top technology companies listed in the NASDAQ100 index. Similarly, we define the rest of the economy to be the remaining non-technology companies in the NASDAQ100 index. More technical details on this are illustrated in the remaining part of this section. 

\subsubsection{Model}
After all the pre-processing and making our series stationary, we are now ready to apply the standard functions in time series. Note that without making series stationary, we were not able to use this model. Since all the pre-processing is already done, this section is going to be very short one for us. We introduce the model and then discuss the inferences. \\[2mm]
Our task is to test whether the Tech index series $y_t$ leads the Non-tech index $x_t$ series. Suppose that the series $y_t$ leads the $x_t$ by $k$ time periods. Following time series models of leading indicators as described in Harvey(2020), we basically need to estimate the following equation: 
\begin{equation}
    \Delta x_t = \beta_0 + \beta_1 \Delta y_{t-1} +\beta_2 \Delta y_{t-2} +....+\beta_p \Delta y_{t-p} + \epsilon_t
\end{equation}
Here, our main coefficient of interest is the vector of $\beta$s. If this vector $\beta$ is significantly different from zero then we can say that the technology index leads the non-technology index.  Therefore, we can formulate the null hypothesis as: 
\begin{align*}
    H_0 : & \quad \beta_1=\beta_2=...=\beta_p=0 
\end{align*}
Note that we cannot directly use $y_t$ or $x_t$ in model because these series are not stationary and therefore does not satisfy the assumption needed to use the model. Therefore we use the first differenced series $\Delta y_t$ and $\Delta x_t$
because they are stationary series.
\subsubsection{Result and Inferences}
For simplicity, we used the number of lag to be $p=10$, however we checked for multiple value of $p$, the qualitative results remain the same. 
The F-statistics of this hypothesis test is $F(10,4252)=1096$, which gives a p-value much lower than 0.001. This means that we can conclude that there is good enough evidence to conclude that the Tech index (series $y_t$) can be a leading indicator for the Non-tech index (series $x_t$). 

One should be careful in interpreting our results, we are saying that $y_t$ can be an indicator to lead $x_t$, that is what will happen in $x_t$ in the future, we can tell by what is happening in $y_t$ today (and some previous periods). However, this relationship cannot be claimed as causal. As they say \textit{correlation does not imply causation}, one may not interpret these results as $y_t$ causing the future $x_t$. Being a leading indicator is a much weaker condition, it can be established without causation. We also checked whether a reverse relation exists i.e. whether $x_t$ can be a leading variable for $y_t$, and we found that there is indeed a reverse relationship. Therefore, establishing causality will be a very difficult challenge in this context. 

\subsection{External Validity of the Results}

\subsubsection{The Problem}
 One can raise the question that whether our analysis is valid for the entire economy. This question basically arises from the fact that we are considering a handful number of companies included in the NASDAQ100 index. There are only 100 total companies in this index and out of these 100, both the technology sector index ad the non-technology sector index is constructed. The point is that there may be more companies out there in the technology sector which are not included in NASDAQ100. Similarly, there could be other non-technology companies outside NASDAQ100. What is happening to those companies doesn't appear in our tech and non-tech indices. Therefore, in some sense what we conclude here in this paper may not be true for the entire technology sector. It could just be true for NASDAQ100 companies which at best are a subset of the entire economy.

We have covered a seventeen-year period which essentially covers the entire lifespan of the NASDAQ100's technology index sector. Therefore from a time period point of view, our results essentially cover the entire population.
\subsubsection{Our Response}
An influential paper on the economics of firms Gabaix (2011) says that firm-size distribution follows something known as \textit{power-law} distribution. This distribution says that there are very few companies with very high valuations and a very large number of companies with low valuations. The consequence of this is that broader variation in the economy can be explained by the variation in big companies. Gabaix (2011) shows that ``The idiosyncratic movements of the largest 100 firms in the United States appear to explain about one-third of variations in output growth''. Since NASDAQ100 contains top-100 companies, therefore, we can say that our findings can well be valid for the entire economy. 
The basic logic is a large portion of the value of a particular sector is contributed by top companies. An alternate response could be to include more companies. One can consider Russell2000\footnote{The Russell 2000 Index is a stock market index that tracks the performance of approximately 2,000 small-cap companies in the United States.} companies to see if there is any difference at the small-cap level. However, including more companies brings in idiosyncratic noise as well, therefore the inclusion of more companies may not always be a good idea.

Given Gabaix(2011)'s result, we stick to NASDAQ100 index and have confidence that our results are valid for the entire economy.

\section{Conclusion}
In this project, I investigate the co-movement of technology and non-technology for the last seventeen years. I call their close co-movement `coupling'. Naturally opposite of this phenomenon is decoupling. I show that after the year 2015, the technology and non-technology sectors decouples in terms of their levels and volatility. In addition, I also test whether technology leads the other sectors, \\[2mm]
I learned several lessons. This time series analysis has been a great learning experience for me. I realized how much pre-processing is needed before jumping to the main model. I started by saying that my goal in this exercise is to test decoupling. In order to do this test, we needed to do a lot of pre-processing. When you start an analysis, sometimes, the results guide you to do the next things. In my case, this is at least partially true. When I plotted the two series together, I found that there is an interesting decoupling after 2015 which further widened in 2019.  This result motivated me to consider this project to be a full-fledged research paper later on working on volatility change. 
The key takeaways from this exploratory data analysis (EDA) can be very briefly summarized in two sentences:
First, the technology index series ($y_t$) follows ARIMA(3,1,3) model. On the other hand, the non-technology index follows an ARIMA(2,1,4) model. The I=1 order implies that both the series are non-stationary but their first differenced series becomes stationary. Second, we have enough evidence that the technology index can be said a leading indicator for the remaining sectors of the economy. \\[2mm]
In our main analysis, we conclude that The technology and non-technology sectors are closely coupled for the majority of the time period. If we do a
statistical test of the difference in mean level, then we cannot reject the null that the two series are close in the full sample. But if we decompose the sample into several parts, we can see the decoupling of the two sectors.
Pre-covid period 2015-2020 is the time when the decoupling of technology and non-technology started appearing. Both covid-19 and the global financial crisis do not appear to have a big enough effect on the close co-movement
of the tech and non-tech series. Further in terms of volatility, there is significantly higher volatility in the technology sector index than in the rest of the economy after 2015. We call this volatility decoupling of the technology sector from the rest of the economy. The tech decoupling we are talking about is not originated from covid pandemic effect. However, we note
that Covid-19 magnifies the decoupling.  We further note that post covid period close the volatility gap by increasing the volatility of the non-
technology sectors.  Overall, the period after 2015 is characterized by higher volatility, especially led by technology. Non-technology sector later joins the trend.\\[2mm]
Overall, we do not claim our results to be causal, however, given the number of years and firms they cover, our results represent important insights for the US economy. I'm happy to find several ideas for further research to make causal claims. 

\bibliography{references}
\cite{gabaix2011granular}
\cite{poon2003forecasting}
\cite{hamilton2020time}
\cite{harvey2020time}
\cite{hastie2009elements}
\cite{jacobsen2020statewide}
\cite{romer1994origins}

\end{document}